\documentclass[english]{article}
\usepackage[T1]{fontenc}
\usepackage[latin9]{inputenc}
\usepackage{units}
\usepackage{amsmath}
\usepackage{amssymb}
\usepackage{stmaryrd}

\makeatletter
\newcommand{\lyxaddress}[1]{
	\par {\raggedright #1
	\vspace{1.4em}
	\noindent\par}
}

\makeatother

\usepackage{babel}
\begin{document}
\title{Geometric model for the electron spin correlation}
\author{Ana María Cetto}
\maketitle

\lyxaddress{Instituto de Física, Universidad Nacional Autónoma de México, Mexico}
\begin{abstract}
The quantum formula for the spin correlation of the bipartite singlet
spin state, $C_{Q}(\boldsymbol{a},\boldsymbol{b})$, is derived on
the basis of a probability distribution $\rho(\phi)$ that is generic,
i. e., independent of $(\boldsymbol{a},\boldsymbol{b})$. In line
with a previous result obtained within the framework of the quantum
formalism, the probability space is partitioned according to the sign
of the product $A=\alpha\beta$ of the individual spin projections
$\alpha$ and $\beta$ onto $\boldsymbol{a}$ and $\boldsymbol{b}$.
A specific partitioning and a corresponding set of realizations $\left\{ \phi\right\} $
are associated with every measurement setting $(\boldsymbol{a},\boldsymbol{b})$;
this precludes the transfer of $\alpha$ or $\beta$ from $C_{Q}(\boldsymbol{a},\boldsymbol{b})$
to $C_{Q}(\boldsymbol{a},\boldsymbol{b'})$, for $\boldsymbol{b'}\neq\boldsymbol{b}.$
A geometric model that reproduces the spin correlation serves to validate
our approach, giving a concrete meaning to the quantum result in terms
of a (local random variable) probability dsitribution. 
\end{abstract}

\section{Introduction}

In a recent paper \cite{FOOP21} an analysis was made of the spin
projection operator correlation function $C_{Q}(\boldsymbol{a},\boldsymbol{b})=\left\langle \left(\hat{\boldsymbol{\sigma}}\cdotp\boldsymbol{a}\right)\left(\hat{\boldsymbol{\sigma}}\cdotp\boldsymbol{b}\right)\right\rangle $
for the bipartite singlet spin state. The analysis, conducted strictly
within the framework of the quantum formalism, led to an unequivocal
probabilistic reading. Specifically, the calculation of $C_{Q}(\boldsymbol{a},\boldsymbol{b})$
was shown to entail a partitioning of the probability space, which
is dependent on the directions $(\boldsymbol{a},\boldsymbol{b})$.
This result is the outcome of a purely theoretical analysis; however,
it can be readily translated to the laboratory language, meaning that
the series of values $($$\pm1)$ obtained for the projections $\alpha$
and $\beta$ leading to the experimental correlation $C(\boldsymbol{a},\boldsymbol{b})$,
cannot be mixed or combined with those obtained for $\alpha$ and
$\beta'$ and leading to $C(\boldsymbol{a},\boldsymbol{b}')$, if
$\boldsymbol{b}'\neq\boldsymbol{b}.$

In the present paper we elaborate on the previous results and take
them further to construct the quantum formula for the spin correlation
on the basis of a probability distribution $\rho(\phi)$. The distribution
function is independent of $(\boldsymbol{a},\boldsymbol{b})$; the
dependence on the directions resides exclusively in the subdivision
of the entire probability space into four mutually orthogonal subspaces
and the realization of the set of variables $\left\{ \phi\right\} $
specific to this subdivision. In other words, for a given pair $(\boldsymbol{a},\boldsymbol{b})$,
the entire set $\left\{ \phi\right\} $ is formed by four complementary
subsets $\left\{ \phi\right\} _{ab}^{k}$, leading respectively to
the eigenvalues $A_{k}=\alpha_{k}\beta_{k}$, with $\alpha_{k},\beta_{k}=\pm1$.
Further, given the degeneracy of eigenvalues $A_{k}$, the four probability
subspaces can be merged pairwise to form two mutually exclusive subsets
$\left\{ \phi\right\} _{ab}^{\pm}$, corresponding to $A=\pm1$. A
distribution function $\rho(\phi)$ that reproduces the quantum result
for $C_{Q}(\boldsymbol{a},\boldsymbol{b})$ is obtained on this basis,
and its application is illustrated by means of a specific geometric
model. 

The paper is organized as follows. Section 2 contains a brief introduction
to the quantum description of the bipartite singlet state, followed
by a discussion of the disaggregation of the correlation $C_{Q}(\boldsymbol{a},\boldsymbol{b})$
on the basis of the spin projection eigenfunctions associated with
the directions $(\boldsymbol{a},\boldsymbol{b})$. The correlation
operator is thus expressed in terms of the projection operators in
the product space of the individual spin spaces. In Section 3 a generic
distribution function $\rho(\phi)$ is obtained that reproduces the
quantum spin correlation for the entangled state. A simple geometric
model for the spin orientations serves to give concrete meaning to
the quantum result. 

\section{Quantum description of the bipartite singlet spin correlation}

We consider a system of two $\nicefrac{1}{2}-$spin particles in the
(entangled) singlet state 
\begin{equation}
\left\vert \Psi^{0}\right\rangle =\frac{1}{\sqrt{2}}\left(\left\vert +_{r}\right\rangle \left\vert -_{r}\right\rangle -\left\vert -_{r}\right\rangle \left\vert +_{r}\right\rangle \right),\label{2-2}
\end{equation}
in terms of the simplified (standard) notation $\left\vert \phi\right\rangle \left\vert \chi\right\rangle =\left\vert \phi\right\rangle \otimes\left\vert \chi\right\rangle ,$
with $\left\vert \phi\right\rangle $ a vector in the Hilbert space
of particle 1, and $\left\vert \chi\right\rangle $ a vector in the
Hilbert space of particle 2. The individual state vectors \begin{subequations}
\label{r+-}
\begin{equation}
\left|+_{r}\right\rangle =\cos\frac{\theta_{r}}{2}\left|+_{z}\right\rangle +e^{i\varphi_{r}}\sin\frac{\theta_{r}}{2}\left|-_{z}\right\rangle ,\label{2-4a}
\end{equation}

\begin{equation}
\left|-_{r}\right\rangle =-e^{-i\varphi_{r}}\sin\frac{\theta_{r}}{2}\left|+_{z}\right\rangle +\cos\frac{\theta_{r}}{2}\left|-_{z}\right\rangle ,\label{2-4b}
\end{equation}
\end{subequations} form an orthogonal basis, with $0\leq\theta_{r}\leq\pi$
and $0\leq\varphi_{r}\leq2\pi$, $\theta_{r}$ and $\varphi_{r}$
being the zenithal and azimuthal angles that define the Bloch vector
$\boldsymbol{r}=\mathbf{i}\sin\theta_{r}\cos\varphi_{r}+\mathbf{j}\sin\theta_{r}\sin\varphi_{r}+\mathbf{k}\cos\theta_{r}$
(see, e. g., Ref. \cite{JA}). 

In Eq. (\ref{2-2}) the direction of $\boldsymbol{r}$ is arbitrary,
since the singlet state is spherically symmetric. The projection of
the first spin operator along an arbitrary direction $\boldsymbol{a}$
is described by $(\hat{\boldsymbol{\sigma}}\cdotp\boldsymbol{a})\otimes\mathbb{I}$,
and the projection of the second spin operator along $\boldsymbol{b}$
is described by $\mathbb{I}\otimes(\hat{\boldsymbol{\sigma}}\cdotp\boldsymbol{b})$.
With the purpose of carrying out a detailed calculation of the correlation
\begin{equation}
C_{Q}(\boldsymbol{a},\boldsymbol{b})=\left\langle \Psi^{0}\right|\left(\hat{\boldsymbol{\sigma}}\cdotp\boldsymbol{a}\right)\otimes\left(\hat{\boldsymbol{\sigma}}\cdotp\boldsymbol{b}\right)\left|\Psi^{0}\right\rangle ,\label{2-6}
\end{equation}
we use Eqs. (\ref{r+-}) to obtain \begin{subequations} \label{sigmaa}
\begin{equation}
\left\langle \pm_{r}\right|\hat{\boldsymbol{\sigma}}\cdotp\boldsymbol{a}\left|\pm_{r}\right\rangle =\pm\boldsymbol{r}\cdot\boldsymbol{a}=\pm\cos\theta_{ra}\label{2-8a}
\end{equation}
and 
\[
\left\langle -_{r}\right|\hat{\boldsymbol{\sigma}}\cdotp\boldsymbol{a}\left|+_{r}\right\rangle =\left\langle +_{r}\right|\hat{\boldsymbol{\sigma}}\cdotp\boldsymbol{a}\left|-_{r}\right\rangle ^{*}=e^{i\varphi}(\boldsymbol{\theta}+i\boldsymbol{\varphi})\cdot\boldsymbol{a},
\]
whence
\begin{equation}
\left|\left\langle \mp_{r}\right|\hat{\boldsymbol{\sigma}}\cdotp\boldsymbol{a}\left|\pm_{r}\right\rangle \right|=\mid\boldsymbol{r}\times\boldsymbol{a}\mid.\label{2-8b}
\end{equation}
\end{subequations} In terms of the complete set of vectors in the
composite Hilbert space,
\begin{eqnarray}
\left|\Psi^{1}\right\rangle  & = & \left|+_{r}\right\rangle \left|-_{r}\right\rangle ,\;\;\left|\Psi^{2}\right\rangle =\left|-_{r}\right\rangle \left|+_{r}\right\rangle ,\nonumber \\
\left|\Psi^{3}\right\rangle  & = & \left|+_{r}\right\rangle \left|+_{r}\right\rangle ,\;\;\left|\Psi^{4}\right\rangle =\left|-_{r}\right\rangle \left|-_{r}\right\rangle ,\label{2-10}
\end{eqnarray}
we get, with the help of Eqs. (\ref{sigmaa}),
\begin{equation}
C_{Q}(\boldsymbol{a},\boldsymbol{b})=\left\langle \Psi^{0}\right|\left(\hat{\boldsymbol{\sigma}}\cdotp\boldsymbol{a}\right)\Big(\sum_{k=1}^{4}\left|\Psi^{k}\right\rangle \left\langle \Psi^{k}\right|\Big)\left(\hat{\boldsymbol{\sigma}}\cdotp\boldsymbol{b}\right)\left|\Psi^{0}\right\rangle =\sum_{k=1}^{4}F_{k},\label{2-12}
\end{equation}
with
\[
F_{1}=-\frac{1}{2}(\boldsymbol{r}\cdot\boldsymbol{a})(\boldsymbol{r}\cdot\boldsymbol{b})=F_{2},
\]
\begin{equation}
F_{3}=-\frac{1}{2}\left[(\boldsymbol{r}\times\boldsymbol{a})\cdot(\boldsymbol{r}\times\boldsymbol{b})-i\boldsymbol{r}\cdot(\boldsymbol{a}\times\boldsymbol{b})\right]=F_{4}^{*}.\label{2-14}
\end{equation}

These equations are greatly simplified by making $\boldsymbol{r}$
lie on the plane formed by $\boldsymbol{a}$ and $\boldsymbol{b}$,
i. e., $\varphi_{r}=\varphi_{a}=\varphi_{b}=0$; with $\theta_{ra}=\theta_{r}-\theta_{a}$
and $\theta_{rb}=\theta_{r}-\theta_{b}$ they become
\[
F_{1}=F_{2}=-\frac{1}{2}\cos\theta_{ra}\cos\theta_{rb},
\]

\begin{equation}
F_{3}=F_{4}=-\frac{1}{2}\sin\theta_{ra}\sin\theta_{rb}.\label{2-15}
\end{equation}
The sum of the four terms gives of course $C_{Q}(\boldsymbol{a},\boldsymbol{b})=-\boldsymbol{a}\cdot\boldsymbol{b}.$
The fact that the result depends only on the angle formed by $\boldsymbol{a}$
and $\boldsymbol{b}$ is due to the spherical symmetry of the singlet
spin state. Looking at the terms separately, however, we observe that
$F_{1}+F_{2}$, involving intermediate states ($\left|\Psi^{1}\right\rangle $
and $\left|\Psi^{2}\right\rangle $) of \emph{antiparallel} spins
(along the arbitrary direction $\boldsymbol{r}$), gives the product
of the projections of $\boldsymbol{a}$ and $\boldsymbol{b}$ onto
$\boldsymbol{r}$, whilst $F_{3}+F_{4}$, involving intermediate states
($\left|\Psi^{3}\right\rangle $ and $\left|\Psi^{4}\right\rangle $)
of \emph{parallel} spins, contains their vector products. In other
words, the two spin projection operators $\hat{\boldsymbol{\sigma}}\cdotp\boldsymbol{a}$,
$\hat{\boldsymbol{\sigma}}\cdotp\boldsymbol{b}$ establish a correlation
not just through the intermediate states representing \emph{antiparallel}
spins---as one might naïvely suppose for the entangled spin-zero
state---but also through the intermediate states of \emph{parallel}
spins, $\left|+_{r}\right\rangle \left|+_{r}\right\rangle $ and $\left|-_{r}\right\rangle \left|-_{r}\right\rangle $. 

We now propose an alternative calculation, by resorting to the individual
eigenvalue equations 
\begin{align}
 & \hat{\boldsymbol{\sigma}}\cdotp\boldsymbol{a}\left|\pm_{a}\right\rangle =\alpha\left|\pm_{a}\right\rangle ,\ \alpha=\pm1,\nonumber \\
 & \hat{\boldsymbol{\sigma}}\cdotp\boldsymbol{b}\left|\pm_{b}\right\rangle =\beta\left|\pm_{b}\right\rangle ,\ \beta=\pm1,\label{2-16}
\end{align}
to construct a new orthonormal basis for the bipartite system: 
\[
\left|\phi^{1}\right\rangle _{ab}=\left|+_{a}\right\rangle \left|-_{b}\right\rangle ,\ \ \left|\phi^{2}\right\rangle _{ab}=\left|-_{a}\right\rangle \left|+_{b}\right\rangle ,
\]
\begin{equation}
\left|\phi^{3}\right\rangle _{ab}=\left|+_{a}\right\rangle \left|+_{b}\right\rangle ,\ \ \left|\phi^{4}\right\rangle _{ab}=\left|-_{a}\right\rangle \left|-_{b}\right\rangle ,\label{2-18}
\end{equation}
and write as before
\begin{equation}
C_{Q}(\boldsymbol{a},\boldsymbol{b})=\left\langle \Psi^{0}\right|(\hat{\boldsymbol{\sigma}}\cdotp\boldsymbol{a})\left(\sum_{k=1}^{4}\left|\phi^{k}\right\rangle _{ab}\left\langle \phi^{k}\right|_{ab}\right)(\hat{\boldsymbol{\sigma}}\cdotp\boldsymbol{b})\left|\Psi^{0}\right\rangle .\label{2-20}
\end{equation}
In view of (\ref{2-16}) and (\ref{2-18}), the terms that contribute
to $C_{Q}$ are
\begin{equation}
(\hat{\boldsymbol{\sigma}}\cdotp\boldsymbol{a})\otimes\mathbb{I}\left|\phi^{k}\right\rangle _{ab}\left\langle \phi^{k}\right|_{ab}\mathbb{I}\otimes(\hat{\boldsymbol{\sigma}}\cdotp\boldsymbol{b})=A_{k}\left|\phi^{k}\right\rangle _{ab}\left\langle \phi^{k}\right|_{ab},\label{2-22}
\end{equation}
where 
\begin{equation}
A_{k}=\alpha_{k}\beta{}_{k}\label{2-24}
\end{equation}
are the eigenvalues of the spin correlation operator 
\begin{equation}
\hat{C}_{Q}(\boldsymbol{a},\boldsymbol{b})=\left(\hat{\boldsymbol{\sigma}}\cdotp\boldsymbol{a}\otimes\hat{\boldsymbol{\sigma}}\cdotp\boldsymbol{b}\right)\label{2-23}
\end{equation}
corresponding to the bipartite states $\left\vert \phi^{k}\right\rangle _{ab}$
given according to Eqs. (\ref{2-16}) and (\ref{2-18}) by
\begin{equation}
A_{1}=A_{2}=-1\equiv A^{-},\ A_{3}=A_{4}=+1\equiv A^{+},\label{2-30-1}
\end{equation}
and $\alpha_{k},\beta_{k}$ are the individual eigenvalues corresponding
to $\left|\phi^{k}\right\rangle _{ab}$. Thus from Eqs. (\ref{2-20})
and (\ref{2-22}) we get 
\begin{equation}
C_{Q}(\boldsymbol{a},\boldsymbol{b})=\sum_{k=1}^{4}A_{k}(\boldsymbol{a},\boldsymbol{b})C_{k}(\boldsymbol{a},\boldsymbol{b}),\label{2-26}
\end{equation}
with 
\begin{equation}
C_{k}(\boldsymbol{a},\boldsymbol{b})=|\left(\langle\phi^{k}|_{ab}\right)|\Psi^{0}\rangle|^{2}.\label{2-28}
\end{equation}
It is clear from this expression that the coefficients $C_{k}$ are
nonnegative and add to give
\begin{equation}
\sum_{k=1}^{4}C_{k}(\boldsymbol{a},\boldsymbol{b})=\sum_{k=1}^{4}\langle\Psi^{0}\left(\left|\phi^{k}\right\rangle _{ab}\left\langle \phi^{k}\right|_{ab}\right)|\Psi^{0}\rangle=1.\label{2-32}
\end{equation}

Notice that the operators 
\begin{equation}
\hat{P}^{k}(\boldsymbol{a},\boldsymbol{b})=\left|\phi^{k}\right\rangle _{ab}\left\langle \phi^{k}\right|_{ab}\label{2-25}
\end{equation}
appearing in Eqs. (\ref{2-20}), (\ref{2-22}) and (\ref{2-32}) are
the projection operators in the product space of the individual spin
spaces, $\mathcal{S=\mathcal{S}}_{1}\varotimes\mathcal{S}_{2}$, with
respective eigenvalues given by $A_{k}$. Equation (\ref{2-26}) is
therefore the appropriate spectral decomposition of the spin correlation.
In terms of the projection operators, the spin correlation operator
(\ref{2-23}) takes the form 
\begin{equation}
\hat{C}_{Q}(\boldsymbol{a},\boldsymbol{b})=\sum_{k=1}^{4}A_{k}(\boldsymbol{a},\boldsymbol{b})\hat{P}^{k}(\boldsymbol{a},\boldsymbol{b})\equiv\sum_{k=1}^{4}\hat{C}_{k}(\boldsymbol{a},\boldsymbol{b}),\label{2-27}
\end{equation}
each term in the sum projecting onto one and only one of the four
mutually orthogonal subspaces $\mathcal{U}^{k}(\boldsymbol{a},\boldsymbol{b})$
that add to form space $\mathcal{S}$ \cite{Hass13},
\begin{equation}
S=\mathcal{U}^{1}\oplus\mathcal{U}^{2}\oplus\mathcal{U}^{3}\oplus\mathcal{U}^{4}.\label{2-29}
\end{equation}
In operational terms (\cite{Busch95}, Ch. 2), this means that the
result of every (joint) measurement falls under one and only one of
these (eigen)subspaces. Further, the coefficient $C_{k}$, which in
(\ref{2-26}) appears as the relative weight of eigenvalue $A_{k}$
contributing to the expectation value $C_{Q}(\boldsymbol{a},\boldsymbol{b})$,
is identified with the probability measure, i. e., the probability
of obtaining $A_{k}$ as the result of a measurement, in accordance
with the Born rule (\cite{Khren}, Ch. 1). We have thus completed
the elements used to describe in quantum theory the measurement statistics
obtained through experiment. 

Let us now consider the observable $C_{Q}(\boldsymbol{a},\boldsymbol{b'})$
with $\boldsymbol{b'}\neq\boldsymbol{b}$. The corresponding projection
operators are 
\begin{equation}
\hat{P}^{k}(\boldsymbol{a},\boldsymbol{b'})=\left|\phi^{k}\right\rangle _{ab'}\left\langle \phi^{k}\right|_{ab'},\label{2-34}
\end{equation}
where $\left|\phi^{k}\right\rangle _{ab'}$ is defined as in (\ref{2-18})
with $b$ replaced by $b'$. Therefore, instead of the partitioning
of $\mathcal{S}$ given by (\ref{2-29}) the spectral decomposition
involves now the partitioning into four mutually orthogonal subspaces
$\mathcal{U}^{k}(\boldsymbol{a},\boldsymbol{b'})$, such that every
(joint) measurement falls under one and only one of these subspaces.
In other words, the probability subspaces are specific to the observable
being measured, i. e., to the measurement setting. 

This assigns a clear meaning to the term \emph{measurement dependence}
that has been introduced in the context of the Bell-type inequalities
(see e. g. \cite{Ver13}): according to the present discussion, it
refers to the dependence of the probability subspaces on the measurement
setting. 

\section{Probability distribution for the bipartite singlet spin state}

In order to arrive at a probability distribution for our problem we
need to calculate the coefficients $C_{k}$ given by (\ref{2-28}).
To simplify the calculation one may, without loss of generality, select
the vector $\boldsymbol{r}$ on the plane defined by the directions
$\boldsymbol{a}$ and $\boldsymbol{b}$, so that Eqs. (\ref{r+-})
reduce to
\begin{equation}
\left|+_{r}\right\rangle =\cos\frac{\theta_{r}}{2}\left|+_{z}\right\rangle +\sin\frac{\theta_{r}}{2}\left|-_{z}\right\rangle ,\;\left|-_{r}\right\rangle =-\sin\frac{\theta_{r}}{2}\left|+_{z}\right\rangle +\cos\frac{\theta_{r}}{2}\left|-_{z}\right\rangle .\label{3-2}
\end{equation}
This gives, using Eqs. (\ref{2-2}) and (\ref{2-18}), with $\theta_{ab}=\theta_{a}-\theta_{b}$,
\begin{subequations} \label{Ck}
\begin{equation}
C_{1}(\boldsymbol{a},\boldsymbol{b})=C_{2}(\boldsymbol{a},\boldsymbol{b})=\frac{1}{2}\cos^{2}\frac{\theta_{ab}}{2},\label{3-4a}
\end{equation}

\begin{equation}
C_{3}(\boldsymbol{a},\boldsymbol{b})=C_{4}(\boldsymbol{a},\boldsymbol{b})=\frac{1}{2}\sin^{2}\frac{\theta_{ab}}{2},\label{3-4b}
\end{equation}
\end{subequations} for the relative weights of the four eigenvalues
$A_{k}$ given by (\ref{2-30-1}). Inserted into Eq. (\ref{2-26})
they reproduce the quantum result
\begin{equation}
C_{Q}(\boldsymbol{a},\boldsymbol{b})=-\cos\theta_{ab},\label{3-6}
\end{equation}
as expected. The contributions due to different signs of $\alpha_{k}$
and $\beta_{k}$ contained in $A_{k}=\alpha_{k}\beta_{k}$, pertain
to mutually exclusive, complementary probability subspaces, as discussed
above. 

Let us call $\varPhi$ the entire probability space and $\varPhi_{ab}^{k}$
the four complementary subspaces. Assuming there exists an associated
probability distribution $\rho(\phi)$ that is a function of a continuous
random variable $\phi$ spanning the entire probability space, such
that $\int_{\varPhi}\rho(\phi)d\phi=1$, the contributions to $C_{Q}(\boldsymbol{a},\boldsymbol{b})$
stemming from the four distinct measurement results $A_{k}$ are \begin{subequations}
\label{phik}
\begin{equation}
\int_{\varPhi_{ab}^{1}}\rho(\phi)d\phi=\int_{\varPhi_{ab}^{2}}\rho(\phi)d\phi=\frac{1}{2}\cos^{2}\frac{\theta_{ab}}{2},\label{3-7a}
\end{equation}
\begin{equation}
\int_{\varPhi_{ab}^{3}}\rho(\phi)d\phi=\int_{\varPhi_{ab}^{4}}\rho(\phi)d\phi=\frac{1}{2}\sin^{2}\frac{\theta_{ab}}{2}.\label{3-7b}
\end{equation}
\end{subequations} Alternatively, in view of the degeneracy indicated
in Eq. (\ref{2-30-1}), one may integrate the subspaces $\varPhi_{ab}^{1}$
and $\varPhi_{ab}^{2}$ into a common subspace $\varPhi_{ab}^{-}$,
corresponding to $A^{-}=-1,$ and $\varPhi_{ab}^{3}$ and $\varPhi_{ab}^{4}$
into the complementary subspace $\varPhi_{ab}^{+}$, corresponding
to $A^{+}=+1,$ so that 
\begin{equation}
\int_{\varPhi_{ab}^{-}}\rho(\phi)d\phi=\cos^{2}\frac{\theta_{ab}}{2},\;\int_{\varPhi_{ab}^{-}}\rho(\phi)d\phi=\sin^{2}\frac{\theta_{ab}}{2}.\label{3-8}
\end{equation}
It is essential to note that the distribution $\rho(\phi)$ is the
same function of $\phi$ regardless of the directions $(\boldsymbol{a},\boldsymbol{b})$;
only the separate domains of integration depend on the angle formed
by $\boldsymbol{a}$ and $\boldsymbol{b}$. Changing the measurement
setting (i. e. from $(\boldsymbol{a},\boldsymbol{b})$ to $(\boldsymbol{a},\boldsymbol{b\text{'}})$)
means using a new set of variables $\varPhi$ that is partitioned
accordingly. To make this distinction clear, we denote with $\phi_{ab}$
the variables $\phi$ spanning the complementary probability spaces
$\varPhi_{ab}^{\pm}$, so that
\begin{equation}
C_{Q}(\boldsymbol{a},\boldsymbol{b})=-\int_{\varPhi_{ab}^{-}}\rho(\phi_{ab})d\phi_{ab}+\int_{\varPhi_{ab}^{+}}\rho(\phi_{ab})d\phi_{ab}.\label{3-9}
\end{equation}
It should be stressed that the notation $\phi_{ab}$ does not imply
a \emph{functional} dependence of the random variable $\phi$ on the
measurement setting $(\boldsymbol{a},\boldsymbol{b})$; it is simply
meant to remind us that the \emph{realization} $\phi$ pertains to
the set of realizations carried out under this measurement setting.

\subsection{General probability distribution function}

As noted above, we are looking for a probability distribution function
$\rho(\phi)$ that complies with Eqs. (\ref{3-8}) and therefore reproduces
the quantum correlation (\ref{3-6}). Such a function can be readily
found by observing that \begin{subequations} \label{sinphi}
\[
\cos^{2}\frac{\theta_{ab}}{2}=\frac{1}{2}(1+\cos\theta_{ab})=\frac{1}{2}\int_{\theta_{ab}}^{\pi}\sin\phi d\phi,
\]
\[
\sin^{2}\frac{\theta_{ab}}{2}=\frac{1}{2}(1-\cos\theta_{ab})=\frac{1}{2}\int_{0}^{\theta_{ab}}\sin\phi d\phi.
\]
\end{subequations} Therefore, the distribution function 
\begin{equation}
\rho(\phi)=\frac{1}{2}\sin\phi,\;0\leq\phi\leq\pi\label{3-12}
\end{equation}
 is a general solution to our problem. With Eq. (\ref{3-12}) the
quantum correlation (\ref{3-9}) is given by
\begin{equation}
C_{Q}(\boldsymbol{a},\boldsymbol{b})=\left(\int_{0}^{\theta_{ab}}-\int_{\theta_{ab}}^{\pi}\right)\rho(\phi_{ab})d\phi_{ab}=-\cos\theta_{ab},\label{3-14}
\end{equation}
where the notation $\phi_{ab}$ reminds us that $\phi$ pertains to
the set of realizations carried out under the measurement setting
$(\boldsymbol{a},\boldsymbol{b})$. It is interesting to note that
the same formula for the distribution, Eq. (\ref{3-12}), has been
previously obtained by Oaknin (\cite{Oak16}, see also \cite{Oak20}),
also within the standard framework of quantum mechanics. By giving
up the assumption implicit in the proof of Bell's inequalities that
there exists an absolute reference frame of angular coordinates for
the entagled bipartite system, Oaknin concludes that the probability
distribution is necessarily given by a function of the form of Eq.
(\ref{3-12}). The variable of integration can of course be changed
to $x_{ab}=\cos\phi_{ab}$ ($-1\leq x_{ab}\leq1$), in which case
$\rho(x_{ab})=\frac{1}{2}$ and Eq. (\ref{3-14}) becomes
\begin{equation}
C_{Q}(\boldsymbol{a},\boldsymbol{b})=\frac{1}{2}\left(\int_{\cos\theta_{ab}}^{1}-\int_{-1}^{\cos\theta_{ab}}\right)dx=-\cos\theta_{ab}.\label{3-15}
\end{equation}

\subsection{A geometric model for the spin correlation\label{model}}

Given that we have found a general probability distribution and an
appropriate separation of the probability space that accounts for
the positive and negative outcomes contributing to the spin correlation,
we now explore a possible geometric explanation for this result. 

With this purpose in mind, let us take a pair of entangled spins and
consider the situation in which the sign of the projection of spin
1 onto $\boldsymbol{a}$ has been determined, say $\alpha=+1$; for
simplicity in the discussion take the $+z$ axis along the direction
$\boldsymbol{a}$, and the $x$ axis perpendicular to it. If the bipartite
system is in the singlet state, we know for sure that the projection
of spin 2 onto the $+z$ axis would give -1. This means that spin
2 lies in the lower half plane, forming any angle $\phi$ such that
$0\leq\phi\leq\pi$, with the origin of $\phi$ along the $-x$ axis
and $\phi$ increasing counterclockwise. Conversely, if the sign of
the projection of spin 1 is $\alpha=-1$, the second spin lies in
the upper half plane, forming any angle $\phi$ such that $0\leq\phi\leq\pi$,
with the origin of $\phi$ along the $x$ axis. In both cases, $A=-1$.
(The argument is of course reversible, in the sense that the sign
of the projection of spin 2 can be defined first, in which case the
angle variable $\phi$ refers to spin 1.) In summary, any series of
measurements along parallel directions gives perfect anticorrelation,
$C_{Q}(\boldsymbol{a},\boldsymbol{a})=C_{Q}(\boldsymbol{b},\boldsymbol{b})=-1$.

Consider now a series of measurements carried out to determine the
correlation of the spin projections onto directions $(\boldsymbol{a},\boldsymbol{b})$
with the $+z$ axis again along $\boldsymbol{a}$, and $\boldsymbol{b}\neq\boldsymbol{a}$.
Take first the case $\alpha=+1$ for spin 1: when spin 2, lying in
the lower half plane, is projected onto the direction $\boldsymbol{b}$
forming an angle $\theta_{ab}$ with the $+z$ axis, $A$ will still
be negative for any angle $\phi$ such that $\theta_{ab}\leq\phi\leq\pi$,
whilst it will become positive for $0\leq\phi\leq\theta_{ab}$. This
gives a concrete meaning to Eq. (\ref{3-14}). What is it that determines
in each instance the specific value of the (random) variable $\phi$
is unknown; we only know its probability distribution. 

When the direction $\boldsymbol{b}$ is changed to $\boldsymbol{b'}$,
a different series of measurements is carried out, with the probability
space subdivided accordingly:

\begin{equation}
C_{Q}(\boldsymbol{a},\boldsymbol{b'})=\left(\int_{0}^{\theta_{ab'}}-\int_{\theta_{ab'}}^{\pi}\right)\rho(\phi_{ab'})d\phi_{ab'}=-\cos\theta_{ab'}.\label{3-14-1}
\end{equation}
The subdivision depends on the range of values of the random variable
$\phi_{ab}$ for which the sign of the product $A=\alpha\beta$ is
either positive or negative. This means that neither $\alpha$ nor
$\beta$ may be transferred from (\ref{3-14}) to (\ref{3-14-1});
not even if the direction of $\boldsymbol{a}$ remains fixed. Precisely
herein lies the essence of the correlation.

Incidentally, a similar reasoning can be applied to the spin correlation
for a single electron, 
\begin{equation}
C(\boldsymbol{a},\boldsymbol{b})=\left\langle \psi\right|\left(\hat{\boldsymbol{\sigma}}\cdotp\boldsymbol{a}\right)\left(\hat{\boldsymbol{\sigma}}\cdotp\boldsymbol{b}\right)\left|\psi\right\rangle .\label{3-16}
\end{equation}
In this case, when the spin projection onto $\boldsymbol{a}$ (taken
again along the $+z$ axis) is +1, its projection onto $\boldsymbol{b}$
is $+1$ (i. e., $A=+1)$ for any angle $\phi$ such that $\theta_{ab}\leq\phi\leq\pi$,
whilst it is $\text{-1}$ (i. e., $A=-1)$ for $0\leq\phi\leq\theta_{ab}$,
and inversely if the spin projection onto $\boldsymbol{a}$ is negative.
The two contributions taken together give the quantum result
\begin{equation}
C(\boldsymbol{a},\boldsymbol{b})=\left(-\int_{0}^{\theta_{ab}}+\int_{\theta_{ab}}^{\pi}\right)\rho(\phi_{ab})d\phi_{ab}=\cos\theta_{ab},\label{3-18}
\end{equation}
with $\rho(\phi)$ given by (\ref{3-12}). We observe that in the
one-particle case the first measurement (say onto $\boldsymbol{a}$
along the $z$ direction) is equivalent to a preparation of the system
for a measurement of the second projection onto $\boldsymbol{b}$.
In the bipartite case the measurement of the two spin projections
counts as a single event (i. e., it is a joint measurement); yet having
chosen the result of the projection of spin 1 (say onto $\boldsymbol{a}$
along the $z$ direction) can be considered equivalent to a 'preparation'.
In both cases illustrated here, $\rho(\phi_{ab})$ plays the role
of a probability density conditioned by the outcome of the projection
onto $\boldsymbol{a}$. 

\textbf{Acknowledgments}. The author is grateful to David Oaknin for
drawing her attention to Refs. \cite{Oak16} and \cite{Oak20}. Valuable
suggestions from an anonymous referee are gratefully acknowledged.

\end{document}